\documentclass[preprint2]{aastex62}

\newcommand{\secref}[1]{\S\ref{#1}}

\shorttitle{}
\shortauthors{Kim et al.}

\begin{document}

\title{Gemini Multi-Object Spectrograph Integral Field Unit Spectroscopy of \\the Double-peaked Broad Emission Line of a Red Active Galactic Nucleus\footnote{dh.dr2kim@gmail.com (DK); mim@astro.snu.ac.kr (MI)}}
	
\author[0000-0002-6925-4821]{Dohyeong Kim}
\altaffiliation{KIAA Fellow}
\affiliation{Kavli Institute for Astronomy and Astrophysics, Peking University, Beijing 100871, China}
\affiliation{Center for the Exploration of the Origin of the Universe (CEOU), Astronomy Program, Department of Physics and Astronomy, Seoul National University, Seoul 151-742, Republic of Korea}
\affiliation{Astronomy Program, Department of Physics and Astronomy, Seoul National University, Seoul 151-742, Republic of Korea}

\author[0000-0002-8537-6714]{Myungshin Im}
\affiliation{Center for the Exploration of the Origin of the Universe (CEOU), Astronomy Program, Department of Physics and Astronomy, Seoul National University, Seoul 151-742, Republic of Korea}
\affiliation{Astronomy Program, Department of Physics and Astronomy, Seoul National University, Seoul 151-742, Republic of Korea}
	
\author[0000-0002-3560-0781]{Minjin Kim}
\affiliation{Department of Astronomy and Atmospheric Sciences, Kyungpook National University, Daegu 702-701, Republic of Korea}
\affiliation{Korea Astronomy and Space Science Institute, Daejeon 305-348, Republic of Korea}
	
\author[0000-0001-6947-5846]{Luis C. Ho}
\affiliation{Kavli Institute for Astronomy and Astrophysics, Peking University, Beijing 100871, China}
\affiliation{Department of Astronomy, School of Physics, Peking University, Beijing 100871, China}	
	
\begin{abstract}
 Galaxy mergers are expected to produce multiple supermassive black holes (SMBHs) in close-separation,
but the detection of such SMBHs has been difficult.
2MASS J165939.7$+$183436 is a red active galactic nucleus (AGN) that
is a prospective merging SMBH candidate owing to
its merging features in Hubble Space Telescope imaging and double-peaked broad emission lines (BELs).
Herein, we report a Gemini Multi-Object Spectrograph Integral Field Unit observation of
a double-peaked broad H$\alpha$ line of 2MASS J165939.7$+$183436. 
Furthermore, we confirm the existence of two BEL peaks that are kinematically separated by 3000\,$\rm km\,s^{-1}$,
with the SMBH of each BEL component weighing at $10^{8.92\pm0.06}\,M_{\rm \odot}$ and $10^{7.13\pm0.06}\,M_{\rm \odot}$,
if they arise from independent BELs near the two SMBHs.
The BEL components were not separated at $>0\farcs1$;
however, under several plausible assumptions regarding the fitting of each spaxel,
the two components are found to be spatially separated at $0\farcs085$ ($\sim250$\,pc).
Different assumptions for the fitting can lead to
a null ($< 0\farcs05$) or a larger spatial separation ($\sim0\farcs15$).
Given the uncertainty regarding the spatial separation, various models, such as
the disk emitter and multiple SMBH models, are viable solutions to explain the double BEL components.
These results will promote future research for finding more multiple SMBH systems in red AGNs,
and higher-resolution imaging validates these different models.

\end{abstract}
	

\section{Introduction} \label{sec:intro}
 According to $\Lambda$ cold dark matter ($\Lambda$CDM) cosmology,
massive galaxies grow via galaxy merging events.
Since almost all spheroidal galaxies harbor SMBHs at centers of their bulges \citep{kormendy13},
multiple SMBHs are expected to exist in the core of merging galaxies,
with some of them merging to form a highly massive SMBH.
Therefore, finding multiple SMBH systems in the process of merging or in close-separation is crucial for
understanding the hierarchical growth of galaxies and the growth of SMBHs.

 Several of the multiple SMBH systems are expected to be binary SMBHs (bSMBHs) or recoiling SMBHs (rSMBHs) after a bSMBH merger.
rSMBHs are merged SMBHs recoiling at a speed of up to several thousand $\rm km\,s^{-1}$ (e.g., \citealt{campanelli07})
owing to anisotropic gravitational wave (GW) emission at the time of merging \citep{peres62}.
Moreover, there have been many efforts to detect bSMBHs and rSMBHs.
To date, various observational features have been suggested to indicate the presence of bSMBHs:
(i) spatially separated cores (e.g., \citealt{myers08,green10});
(ii) double-peaked BELs (e.g., \citealt{boroson09}); and
(iii) double-peaked narrow emission lines (NELs; e.g., \citealt{woo14}).

To find rSMBHs, previous studies have used two observational characteristics:
(i) the spatial offset between the centers of the broad line region (BLR) and the host galaxy (e.g., \citealt{lena14}); and
(ii) BELs having a velocity offset with respect to the systemic velocity (e.g., \citealt{komossa08}).
Although systems exhibiting the aforementioned properties have been detected,
such claims are not yet fully convincing (e.g., \citealt{shields09,chornock10,condon17}).

\begin{figure*}
	\centering
	\includegraphics[width=\textwidth]{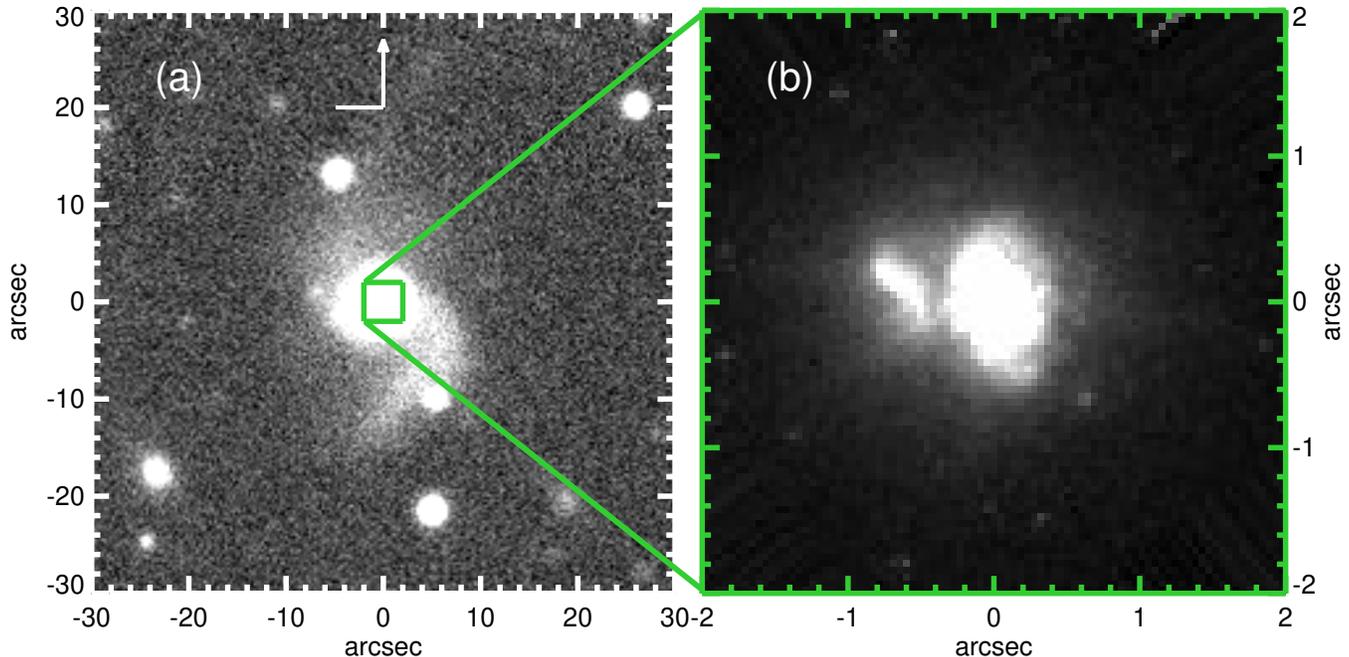}\\
	\caption{(a) DECaLS $r$ band and (b) HST WFPC2 $F814W$ images of 1659$+$1834. 
		The DECaLS image shows extended low surface brightness features that
		can be interpreted as tidal tails resulting from a galaxy merger. 
		The white arrow denotes north.
		The HST image shows the central structures of the inner region.\label{fig:HST_Image}}
\end{figure*}

 Some of the possible candidates that are
well-suited for detecting bSMBHs and rSMBHs are red active galactic nuclei (AGNs).
Red AGNs are AGNs that exhibit a highly red colors owing to
dust obscuration (e.g., $J$--$K>1.7$, $R$--$K> 4.0$, and $E(B$--$V) > 0.1$ in \citealt{glikman07}).
Moreover, based on simulations, red AGNs are considered to be the final stage of a galaxy merger \citep{hopkins08,blecha18};
in addition, this consideration is based on observational grounds such as
(i) dusty red colors \citep{kim18a}, (ii) high BH accretion rates \citep{kim15b,kim18b,kim18a},
(iii) enhanced star formation activities \citep{georgakakis09},
(iv) high fractions of merging features \citep{urrutia08,glikman15}, and
(v) young radio jets \citep{georgakakis12}.

 Recently, we identified a red AGN, 2MASS J165939.7$+$183436 (hereafter 1659$+$1834),
that has a double-peaked BEL profile and a host galaxy with a merging feature.
Here, we present a spatially resolved spectroscopic observation of the core of 1659$+$1834
using the Gemini Multi-Object Spectrograph (GMOS) Integral Field Unit (IFU)
as an attempt to find the separation of the two BELs at $\gtrsim 0\farcs1$.

 In this study, we adopt a standard $\Lambda$CDM model of
$H_{0}=70\,{\rm km\,s^{-1}}$\,Mpc$^{-1}$, $\Omega_{m}=0.3$, and $\Omega_{\Lambda}=0.7$, 
which is supported by observational studies conducted in the past decades
(e.g., \citealt{im97,planck16,planck18}).
Therefore, 1 arcsec corresponds to $\sim$2.9\,kpc at a redshift of 1659$+$1834, $z=0.170$.

\section{1659$+$1834} \label{sec:sample}
 Our target, 1659$+$1834, is a nearby ($z=0.170$) bright (12.9 mag in the $K_{\rm s}$ band) AGN 
and one of the 2MASS-selected red AGNs \citep{marble03}.
The 2MASS-selected red AGNs
are spectroscopically confirmed quasars with
$J$--$K_{\rm s} > 2$ \citep{cutri01,smith02}.
Not all of the AGNs selected using this NIR color criterion are dust-obscured red AGNs \citep{kim18b},
but 1659$+$1834 is a dust-obscured red AGN with $E(B$--$V) \sim 0.5$ \citep{kim18b}.

\begin{figure*}
	\centering
	\includegraphics[width=\textwidth]{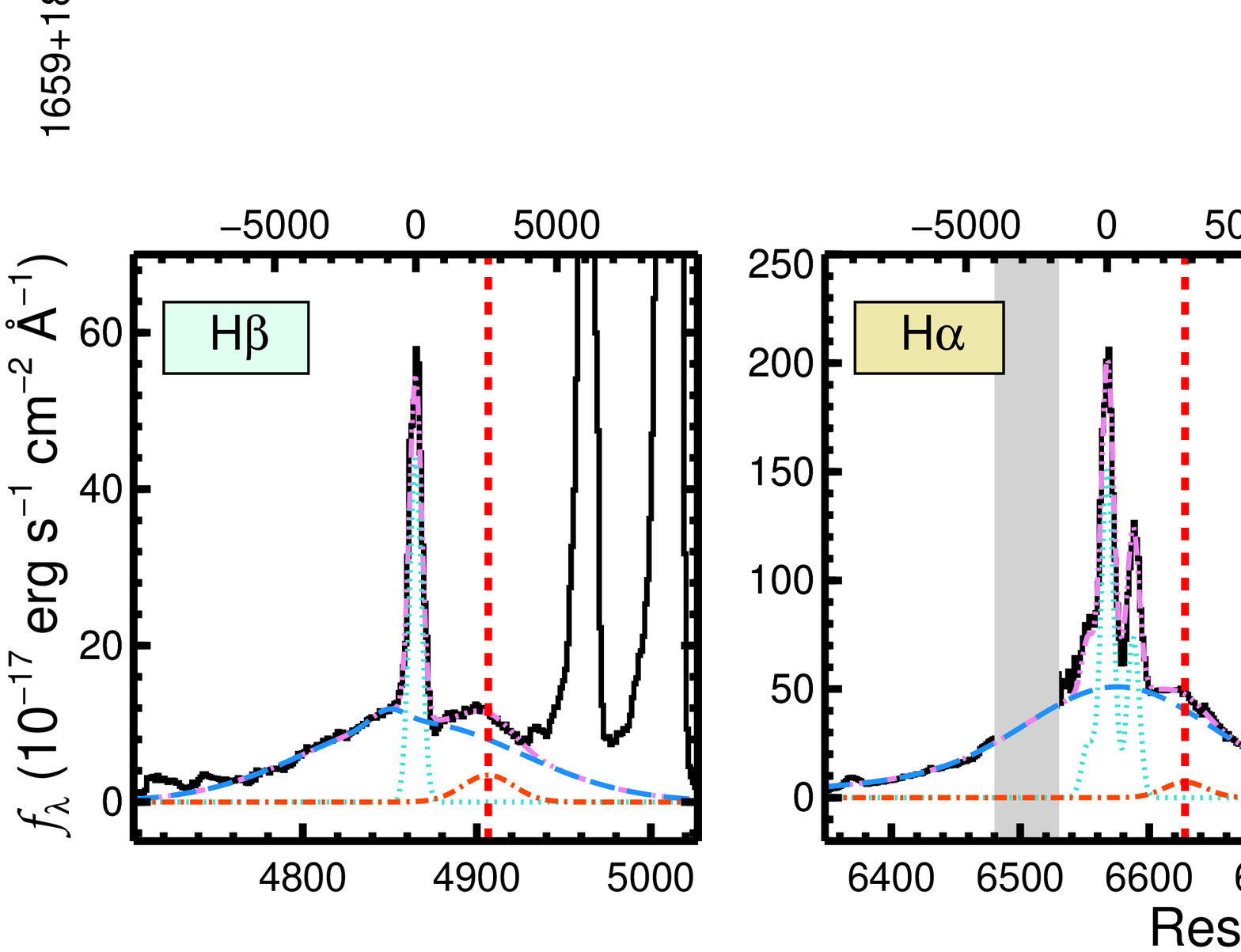}\\
	\caption{H$\beta$, H$\alpha$, P$\beta$, and P$\alpha$ lines of 1659$+$1834.
		The black lines are the continuum-subtracted spectrum sourced from \cite{kim18b},
		and the purple lines represent the best-fit models.
		The cyan lines indicate the narrow lines, and the blue and red lines represent the primary and secondary components, respectively.
		The red vertical dashed lines represent the central wavelengths of secondary components.
		In the H$\alpha$ line, the gray box represents the spectral region affected by telluric lines.\label{fig:Lines}}
\end{figure*}

 Similar to several other dust-obscured red AGNs,
1659$+$1834 exhibits the morphology of a galaxy merger.
Figure \ref{fig:HST_Image} shows the surrounding region and the central part of 1659$+$1834.
The larger-field-of-view image
from the Dark Energy Camera Legacy Survey (DECaLS; \citealt{dey19}) reveals tidal tails,
and Hubble Space Telescope (HST) $\it{F814W}$ imaging
\footnote{Based on observations made with the NASA/ESA Hubble Space Telescope,
	obtained from the data archive at the Space Telescope Science Institute.
	STScI is operated by the Association of Universities for Research in Astronomy
	Inc. under NASA contract NAS 5-26555.}
of the central part shows two galaxy components,
which support the merger hypothesis.

 Since 1659$+$1834 was spectroscopically observed in the optical wavelength range in 1999 \citep{smith00},
several other spectroscopic observations have been made in the optical range in 2004,
for example, through the Sloan Digital Sky Survey (SDSS) and
\cite{canalizo12} using the Echellette Spectrograph and Imager (ESI; \citealt{sheinis02}).
In addition, an NIR spectrum was obtained in 2016 \citep{kim18b}.

 Figure \ref{fig:Lines} shows the H$\beta$, H$\alpha$, P$\beta$, and P$\alpha$ lines of 1659$+$1834 obtained from 
the optical and NIR spectra presented in \cite{kim18b}
and the result of spectral fitting using the method described in \secref{sec:analysis}.
The figure demonstrates that 1659$+$1834 has double-peaked BELs comprising big blue and small red components.
We define the big blue and small red components as the primary and secondary components, respectively.
Only the primary components have the corresponding NELs.

 The FWHM values of
the secondary components are $\sim$2100\,$\rm km\,s^{-1}$,
and those of the primary components are $\sim$4000--8500\,$\rm km\,s^{-1}$
(e.g., H$\beta$: 7980$\pm$260, H$\alpha$: 8490$\pm$90, P$\beta$: 5310$\pm$170, and P$\alpha$: 4090$\pm$170\,$\rm km\,s^{-1}$).
The secondary BEL components are redshifted by $\sim 3000\,{\rm km\,s^{-1}}$ from the NELs,
as shown in Figure \ref{fig:Lines}.

 Figure \ref{fig:Lines} shows that the line-flux ratio trends of the primary and secondary components are different.
The observed flux ratios of P$\alpha$/H$\beta$, P$\beta$/H$\beta$, and H$\alpha$/H$\beta$ are
1.43$\pm$0.38, 1.58$\pm$0.33, and 6.96$\pm$1.36 for the primary component, and
2.45$\pm$0.31, 1.55$\pm$0.11, and 2.32$\pm$0.88 for the secondary component.
The different line ratios of the primary and secondary components suggest that
the two components may arise from physically different BLRs.
However, the large errors in some of the line ratio measurements make such a claim inconclusive.
Notably, after internal extinction correction ($E(B-V)=0.505$; \citealt{kim18b}),
the line-flux ratios of the primary component are consistent with the general line-flux ratios of type 1 AGNs
(e.g., $L_{\rm P\alpha}$/$L_{\rm H\beta}$: 0.37, $L_{\rm P\beta}$/$L_{\rm H\beta}$: 0.34,
and $L_{\rm H\alpha}$/$L_{\rm H\beta}$: 3.1; \citealt{dong08,kim10}).

 When the spectra obtained in 1999 \citep{smith00}, 2004 \citep{canalizo12}, and 2018 (this work)
are compared, no change is observed in the central velocity offset over
the 20 yr-period between the primary and secondary components at $\Delta V < 50\,{\rm km\,s^{-1}}$.
The constant velocity offset over 20 yr suggests that
the physical separation between the two components is $\gtrsim$1 pc,\footnote{This limit is calculated by
assuming that the secondary component is orbiting around the primary component in a circular orbit
with an orbital speed of ${\rm 3000\,km\,s^{-1}}$,
where we expect to observe the velocity offset change due to the change in the radial velocity component along the line of sight}
if one component orbits around the other component.

\section{Observation and Data Reduction} \label{sec:observation}
 Using the GMOS IFU on the Gemini North telescope \citep{allington02},
we obtained the spatially resolved spectra of 1659$+$1834 on 2018 May 20.
The data were obtained under clear weather,
and the seeing, determined from the guide star image,
ranged from 0$\farcs$3 to 0$\farcs$6.

 The spectra were obtained using a one-slit-mode setup with R400 grating,
which provides a spectral resolution of $R \sim 3100$ across 4500--8800\,$\rm \AA{}$.
This observational setup provides a field of view (FoV) of 5$\farcs$0$\times$3$\farcs$5,
which is sufficiently wide to include the central structures of 1659$+$1834.
This FoV is sampled using hexagonally shaped lenslets with a radius of 0$\farcs$1 \citep{allington02}.

 The observation was performed using 12 exposures, each lasting for 1200\,s; hence, the total integration time was 14,400\,s.
For the flux calibration, we observed a nearby standard star after the sample observation using the same observational setup.

 The spectra were reduced using the general Gemini IFU reduction package.\footnote{\url{https://gmos-ifu-1-data-reduction-tutorial-gemini-iraf.readthedocs.io}}
Through the reduction process, the data were resampled onto 1700 (34$\times$50) rectangular spaxels with a size of 0$\farcs$1$\times$0$\farcs$1.
The reduction process returns flux variance, which is used to determine the flux uncertainty.
The yielded variance includes the correction for the statistical correlation between adjacent spaxels produced by
resampling large lenslets into fine spaxels (see Section A8 of \citealt{davies15}).

 To estimate errors in the centroid position of a specific component using this procedure,
we compare the spatial distributions of five pairs of two nearby continuum fluxes
((i) 6100--6150\,$\rm \AA{}$ and 6150--6200\,$\rm \AA{}$;
(ii) 6200--6250\,$\rm \AA{}$ and 6250--6300\,$\rm \AA{}$;
(iii) 6850--6900\,$\rm \AA{}$ and 6900--6950\,$\rm \AA{}$;
(iv) 6950--7000\,$\rm \AA{}$ and 7000--7050\,$\rm \AA{}$;
and (v) 7050--7100\,$\rm \AA{}$ and 7100--7150\,$\rm \AA{}$)
assumed to be spatially coincident to each other.
We find that the centroids of the two components are separated by
0.026$\pm$0.008 spaxel without any directional bias,
suggesting that this is the level of accuracy
for the centroid determination with a sufficiently large signal-to-noise ratio (S/N; up to $\sim$75) in the flux.

\begin{deluxetable*}{cccccccc}
	\tablecolumns{8}
	\tablewidth{0pt}
	\tablenum{1}
	\tablecaption{Primary and Secondary Component Properties of the H$\alpha$ Line\label{tab:Ha_prop}}
	\tablehead{
		\multicolumn{3}{c}{Primary Component}& 	\colhead{}&		\multicolumn{4}{c}{Secondary Component}\\
		\cline{1-3} \cline{5-8}\\
		\colhead{$L_{\rm H\alpha}$}&				\colhead{$\rm FWHM_{H\alpha}$}&		\colhead{$M_{\rm BH}$}&			\colhead{}&
		\colhead{$L_{\rm H\alpha}$}&				\colhead{$\rm FWHM_{H\alpha}$}&		\colhead{$M_{\rm BH}$}&			\colhead{Coord.\tablenotemark{1}}\\
		\colhead{($\rm 10^{43}\,erg\,s^{-1}$)}&\colhead{($\rm km\,s^{-1}$)}&	\colhead{($M_{\rm \odot}$)}&			\colhead{}&
		\colhead{($\rm 10^{42}\,erg\,s^{-1}$)}&\colhead{($\rm km\,s^{-1}$)}&	\colhead{($M_{\rm \odot}$)}&			\colhead{(South, East)}
	}
	\startdata
	6.14$\pm$2.02&				6210$\pm$130&		$10^{8.92 \pm 0.06}$&		&
	1.39$\pm$0.48&				1930$\pm$130&		$10^{7.13 \pm 0.06}$&		0$\farcs$068, 0$\farcs$051		
	\enddata
	\tablecomments{	\tablenotemark{1} Coordinate of the secondary component with respect to the primary component}
\end{deluxetable*}

\section{Analysis} \label{sec:analysis}

 To increase the S/N of each wavelength element of each spaxel,
we rebinned the reduced spectrum to have the wavelength pixel resolution of $ \Delta \lambda = \lambda /R$, where $R=3100$.
Typically, 4--5 wavelength elements are combined to form a rebinned wavelength element.
The spectra exhibit high S/Ns of the continuum at 6950\,$\rm \AA{}$,
with the highest S/N of $\sim$75 at the brightest center.

 After shifting the rebinned spectrum to the rest frame at $z=0.170$ \citep{marble03},
the continuum is fitted around the H$\alpha$ line in 218 spaxels that has an optimal S/N ($>$15).
Note that the analysis is restricted to H$\alpha$,
as H$\beta$ did not have a sufficient number of spaxels with an appropriate S/N
to allow the decomposition of the primary and secondary components.
The spatial area covered by these high S/N spaxels corresponds roughly to
the central region within a radius of $\sim$0$\farcs$85.
The seeing condition of 0$\farcs$6 smears the central AGN light to $\sim$1$\farcs$0,
making the AGN light the dominant component in the high S/N spaxels.
Therefore, we adopt a single power-law function to fit the continuum (e.g., \citealt{kim10,kim15a}).
The continuum-fitting regions are chosen as 6100--6250\,$\rm \AA{}$ and 6800--7100\,$\rm \AA{}$ to avoid line features.

 Every fitting procedure is performed using \texttt{MPFIT} \citep{markwardt09}.
\texttt{MPFIT} returns the fitted parameters and their uncertainties, and
the continuum-fitting uncertainties (errors in the fitted parameters)
are added to the residual flux errors in quadrature ($\sigma^2 = \sigma_{\rm fit}^2 + \sigma_{\rm flux}^2$).

\begin{figure}
	\centering
	\includegraphics[width=\columnwidth]{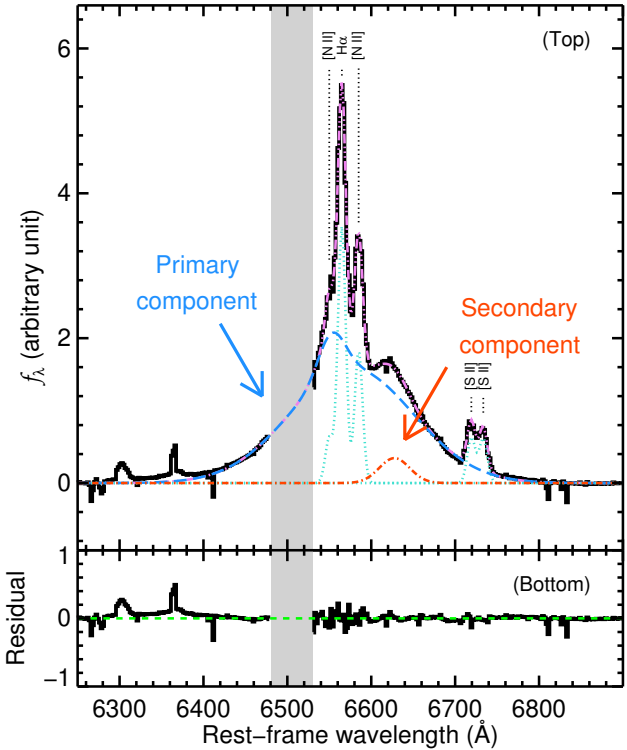}
	\caption{(Top) H$\alpha$ line fitting result in the integrated spectrum.
		The black line indicates the continuum-subtracted spectrum, and
		the gray box represents the masked out spectral region affected by telluric lines.
		The purple line represents the best-fit model, and the cyan line indicates the narrow lines of H$\alpha$ and [\ion{N}{2}] doublets.
		The blue and red lines represent the primary and secondary components, respectively.
		The vertical dotted lines indicate the fitted narrow lines of H$\alpha$, [\ion{N}{2}], and [\ion{S}{2}] doublets.
		(Bottom) The residual of the fitting.
		\label{fig:Ha}}
\end{figure}

 After the continuum subtraction,
we fit the H$\alpha$ line, [\ion{N}{2}] $\lambda\lambda$6548, 6583, and [\ion{S}{2}] $\lambda\lambda$6716, 6731 simultaneously. 
The H$\alpha$ line is decomposed into a narrow line and a double-peaked broad line, as shown in Figure \ref{fig:Ha}.
Note that the wavelength regions of 6250--6280, 6400--6480, 6530-6700, and 6750-6900\,$\rm \AA{}$
are used for the fit, and we omit the spectra at 6280--6400\,$\rm \AA{}$
to avoid the [\ion{O}{1}] $\lambda\lambda$6300, 6363 doublet.

 A single Gaussian function is used for fitting the H$\alpha$ narrow line,
and two double Gaussian functions are assigned for the [\ion{N}{2}] and [\ion{S}{2}] doublets.
These five Gaussian functions are
fitted simultaneously with the same widths, and a fixed flux ratio of 2.96 is applied to the [\ion{N}{2}] doublet \citep{kim06}.
We set free the central wavelength for the H$\alpha$ narrow line,
but those of the other NELs are bound to the H$\alpha$ narrow line.

 NELs are often fitted with double or multiple Gaussian functions (e.g., \citealt{ho97}).
However, the H$\alpha$ narrow line and each line of the [\ion{N}{2}] and [\ion{S}{2}] doublets of 1659$+$1834
are well-fitted with a single Gaussian model \citep{kim18b}
without any significant improvement in the fitting by using a double Gaussian model.
Therefore, we fit the H$\alpha$ narrow line and each line of the [\ion{N}{2}] and [\ion{S}{2}] doublets with a single Gaussian model.

 The H$\alpha$ double-peaked BEL is fitted with three Gaussian functions, and
their widths and central wavelengths are set as free parameters.

 We compare the spatial flux distributions for the three Gaussian functions.
The centers of the distributions of two Gaussian functions are spatially close ($\sim$0.2 spaxel)
at short wavelengths,
but the reddest Gaussian component is $\sim$4 times away from the two blue components.
We group the two short-wavelength Gaussian components into a single component
and name it the primary component.
Furthermore, the longest-wavelength Gaussian component is treated as a separate component called
the secondary component.
During the fit, we fix the flux ratio of the two Gaussian functions of the primary component,
and the fixed ratio is adopted from the measurement in the integrated spectrum obtained from all spaxels.
This procedure is justified if we assume that
the shapes of both of the components do not change at different spaxels
because the line flux of a component originates from a single emitting source.

 We also tried to fit the BEL with four Gaussian components.
However, there was no significant improvement in $\chi^2$, and the added Gaussian function was unreliably broad
($\rm{FWHM} > 10,000\,{\rm km\,s^{-1}}$) with a near-null flux.
Therefore, we used the results obtained using three Gaussian functions.

 The fit provides the FWHM and flux for each spaxel, 
and the FWHM values are adopted after correcting the instrumental spectral resolution,
as $\rm FWHM^2 = FWHM_{\rm obs}^2 - FWHM_{\rm inst}^2$.
\texttt{MPFIT} also yields the fitting uncertainties of the primary and secondary component fluxes,	
which are found to be 4.6\,\% (from 1.1\,\% to 18\,\%)
and 13\,\% (from 3.9\,\% to 49\,\%), respectively, in each of the 218 spaxels.

 Moreover, we measure the FWHM values and luminosities of the primary and secondary components using the integrated spectrum obtained from all spaxels.
The line luminosities are derived assuming that $E(B$--$V)_{\rm line}=0.505\pm0.007$ from line-luminosity ratios \citep{kim18b}.
However, the assumed $E(B$--$V)$ value can be changed to
$E(B$--$V)_{\rm cont} = 0.636\pm0.001$ if the continuum shape is used \citep{kim18b},
and this can cause the line luminosities to vary by 32.9\,\%.
Thus, we consider the 32.9\,\% variation to be an additional uncertainty arising from the extinction correction,
and add this to the flux uncertainties.
The fitting results are listed in Table \ref{tab:Ha_prop}.

\section{Spatial Separation of Two BEL Regions} \label{sec:result}
\begin{figure*}
	\centering
	\includegraphics[width=\textwidth]{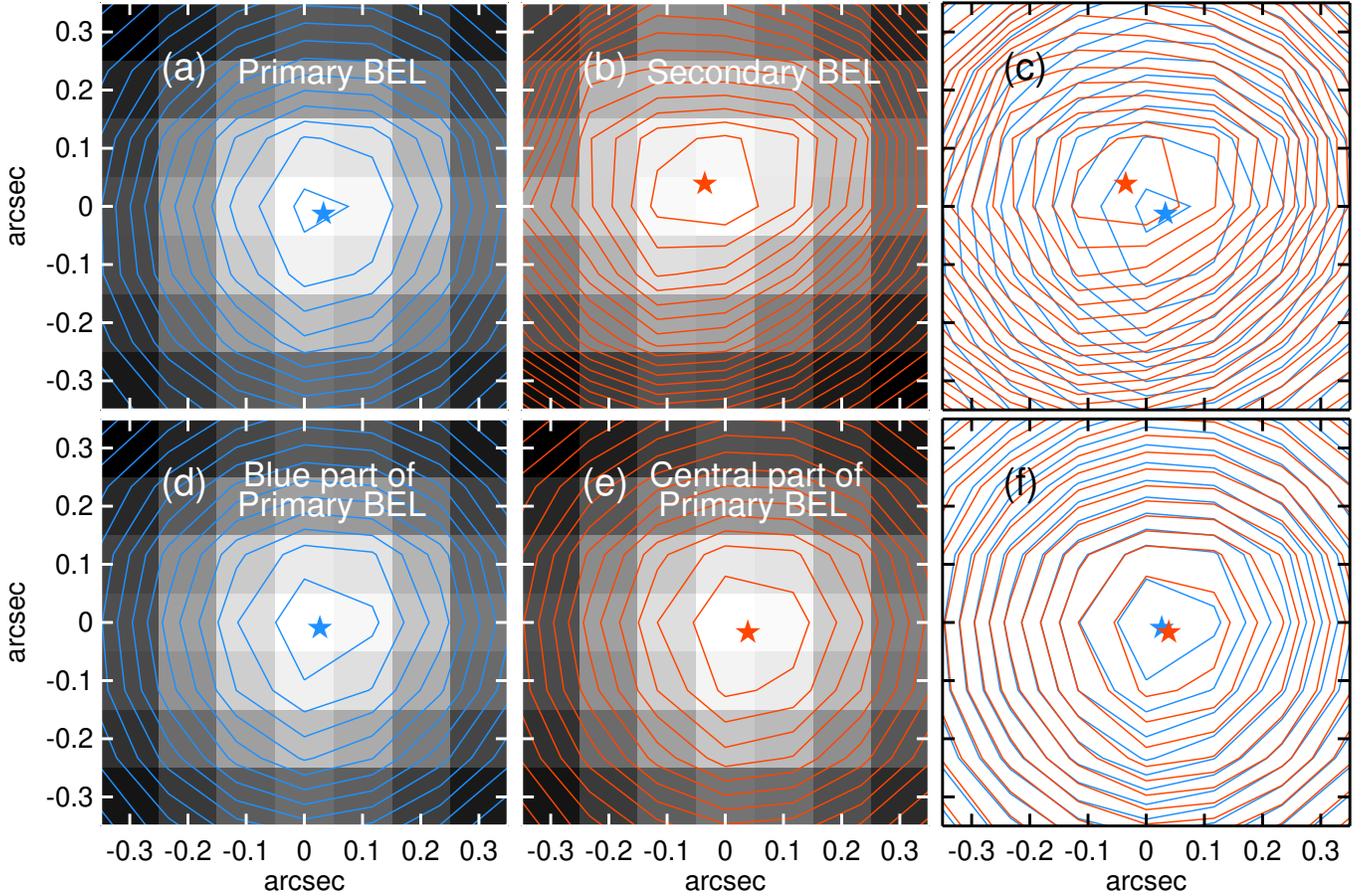}\\
	\caption{
		(a)--(c): Flux images of the primary and secondary BEL components with their contours,
		and their comparison, showing a slight spatial offset between the two components.
		A spatial offset of 0$\farcs$085 is found between (a) and (b).
		(d)--(f): Flux images of the primary BEL components split into two wavelength regions
		to show the expected amount of image shift from a single BEL component.
		Images (d) through (e) show no significant offset.
		\label{fig:Map}}
\end{figure*}

 We construct two flux images for each H$\alpha$ BEL component.
The total flux value of each BEL component obtained from the pixel-by-pixel spectral fit was assigned as the pixel value of the corresponding BEL component.
Panels (a)--(c) of Figure \ref{fig:Map} show the images of the central part (0$\farcs$7$\times$0$\farcs$7) of each component
and a comparison of the primary and secondary components.
We find that the two BEL components originate from the central region,
not from the off-center substructure seen in HST imaging. 
However, the panels of Figure \ref{fig:Map} show that
the secondary component is slightly shifted with respect to the primary component.

 We measure the centroid values of the two distributions via a two-dimensional Gaussian fit.
The fitted centroid values are separated by 0.85 spaxels (0$\farcs$085)
which corresponds to $\sim$250\,pc in a physical scale.

 As a reference, we compare the flux images of
the blue (6420--6480\,$\rm \AA{}$) and central parts (6530--6610\,$\rm \AA{}$) of the H$\alpha$ primary component,
which are shown in panels (d)--(f) of Figure \ref{fig:Map}.
The centroid offset is measured to be 0.13 spaxels, consistent with no spatial offset between the two wavelength images.

\begin{figure}
	\centering
	\includegraphics[width=\columnwidth]{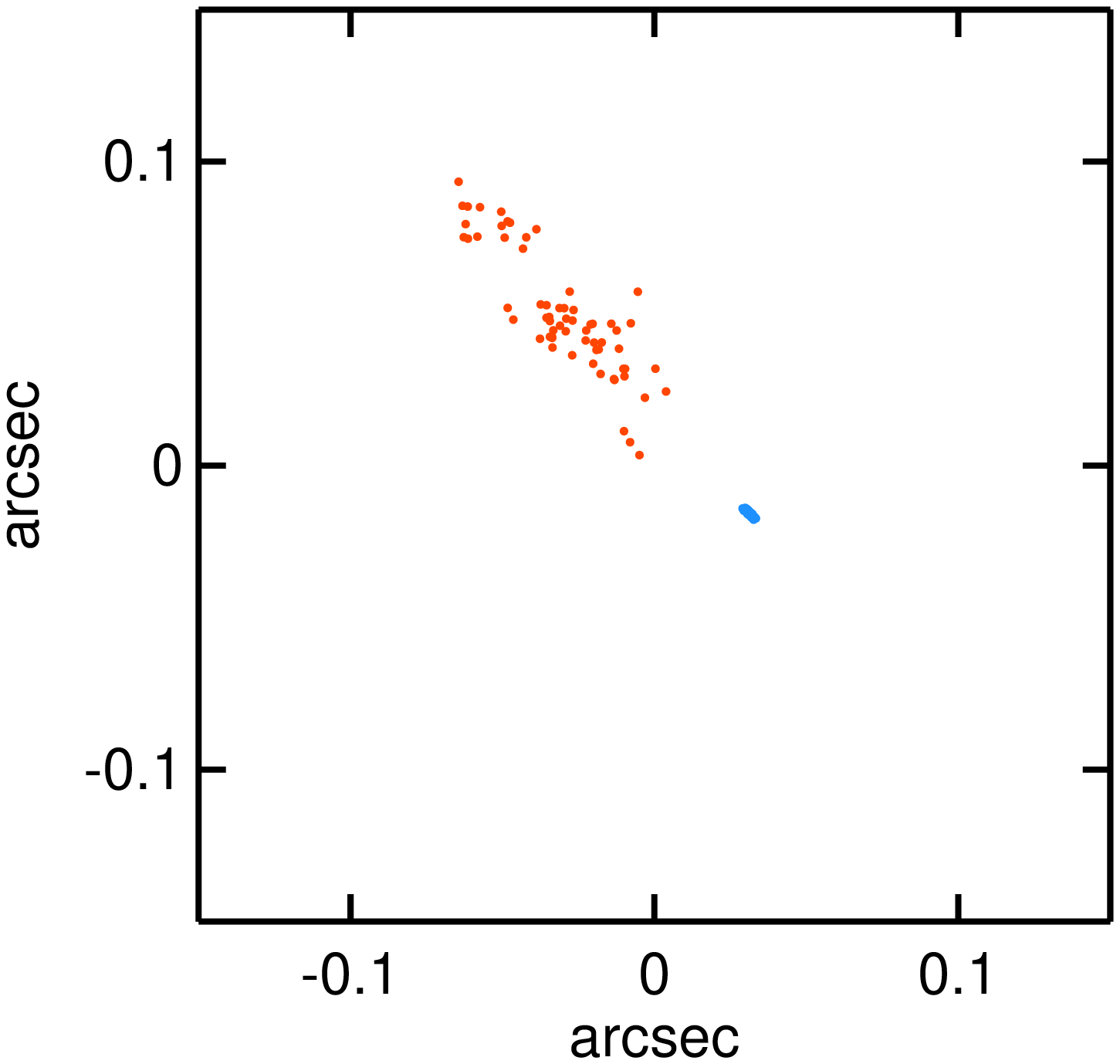}\\
	\caption{ Measured centroids of the primary and secondary BEL components
		for the 64 cases obtained by setting the fitting parameters to free or fixed.
		The blue and red dots indicate the centroids of the primary and secondary components, respectively.
		\label{fig:Par}}
\end{figure}

 However, the measured spatial centroids vary with the manner in which the fitting parameters are fixed.
As a test, we assigned the line widths and central wavelengths of the three Gaussian functions as free or fixed parameters,
where the fixed values are adopted from the measurements in the integrated spectrum.
In total, there are 64 cases for the fit.
The measured centroid offsets vary from 0.39 to 1.48 spaxels
with a median offset of 0.87 spaxels, as shown in Figure \ref{fig:Par}.

 We note that the measured spatial centroid can be changed by setting the fitting condition differently.
If the flux ratios of the three Gaussian components used for the primary and secondary components are set as free,
the measured centroid offset varies from 0.05 to 1.15 spaxels, and the median offset is 0.39 spaxels.
This result implies that the measured centroid offset could be changed depending on the fitting conditions,
and the measured spatial centroids are potentially volatile measurements.

 Moreover, we investigate the effects of the seeing on the separation.
Among the 12 exposures in our observation, the seeing was optimal and stable at 0$\farcs$3--0$\farcs$4 during the first exposure.
From the data obtained from the first exposure, we measure the centroid offset of the two components, which is 0.52 spaxels.
The other 11 exposure data all show offsets of 0.46--1.00 spaxels,
with the median centroid offset of 0.69 spaxels.
This result implies that the separation can only be 0.52 spaxels
and that the measured separation of 0.85 spaxels is exaggerated
because the data were obtained at a suboptimal seeing $\sim$0$\farcs$6.

 The probability of the centroid offset being created by chance is also examined.
To this end, we perform a Monte-Carlo simulation 1000 times by adding H$\alpha$ line flux uncertainties randomly for each spaxel.
However, the uncertainty of the centroid offset arising from the flux uncertainty is only 0.09 spaxels,
which is not significant compared to the other uncertainties.

 Considering these tests, we cannot exclude two possibilities:
First, the separation is smaller or bigger than 0.85 spaxels.
The measured separation can vary from $<$0.5 to 1.5 spaxels depending on the fitting parameter setup.
Second, there is no separation, and the measured separation is a result of several kinds of overlapping uncertainties.
If the separation is $<$0.5 spaxels, the measured separation can arise from the seeing effects,
as a centroid is determined at an accuracy of $\sim$10\,\% of the seeing.

 Otherwise, if the primary and secondary components are indeed spatially separated,
the two components of the BEL arise from independent SMBHs.
We measure the BH masses of these two components using
their respective $L_{\rm H\alpha}$ and $\rm FWHM_{H\alpha}$ values from the integrated spectrum.
The BH masses are estimated by adopting the H$\alpha$-based BH mass estimator (Eq. A1 in \citealt{greene07}).
Thus, we obtained the BH masses of $10^{8.92 \pm 0.06}\,{M_{\odot}}$ and $10^{7.13 \pm 0.06}\,{M_{\odot}}$
for the primary and secondary components, respectively.

\section{Origin of Two BELs} \label{sec:discussion}
\begin{figure*}
	\centering
	\includegraphics[scale=0.20]{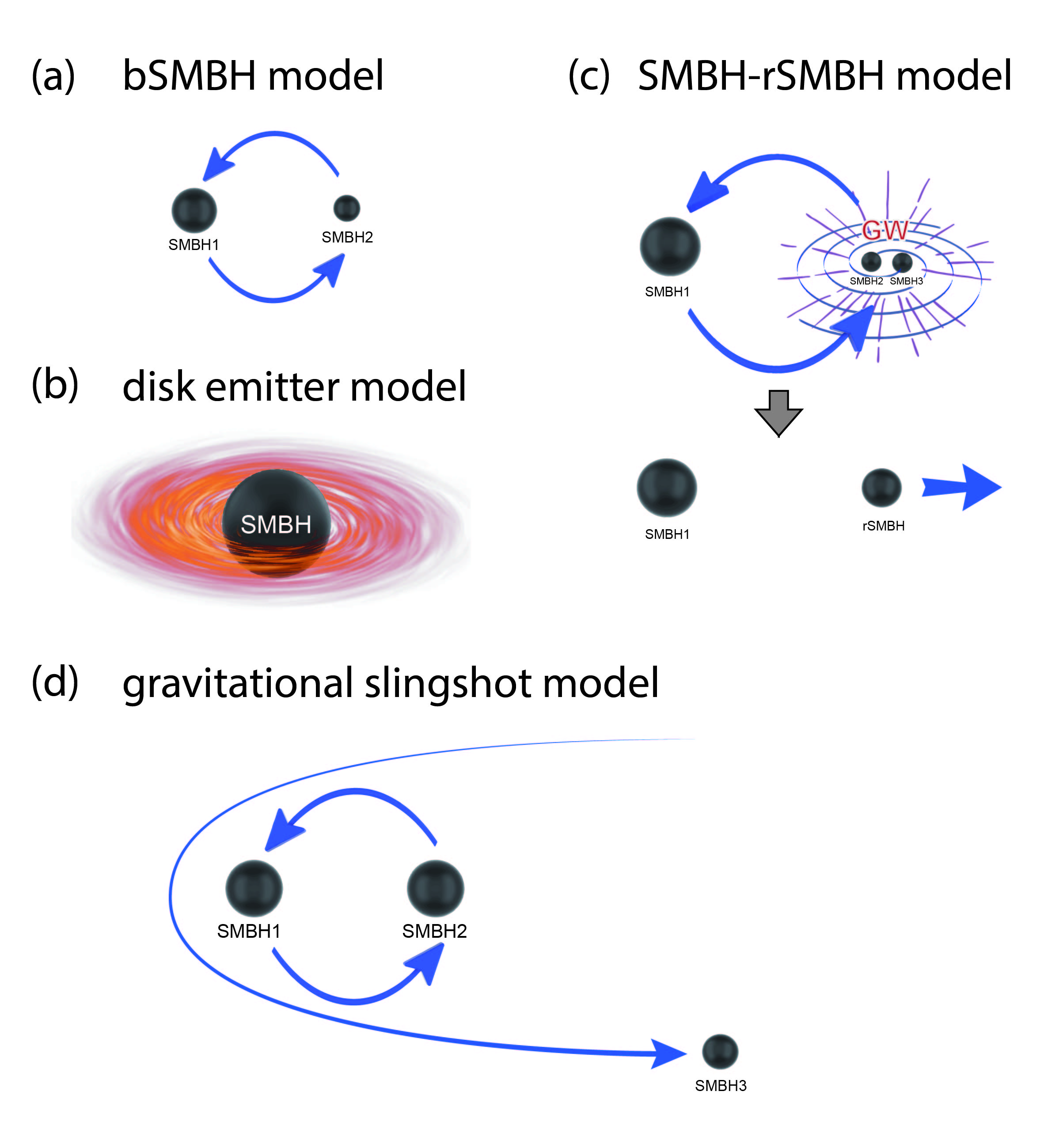}\\ 
	\caption{ Plausible models of the central SMBH system of 1659$+$1834.
		(a) bSMBH model: the bSMBH system is comprised of two SMBHs,
		SMBH1 and SMBH2, and they are orbiting their center of mass.
		The bSMBH model can explain the double-peaked BEL,
		but the measured velocity offset and spatially separated distance of the double-peaked BEL of 1659$+$1834 cannot arise from the bSMBH system.
		(b) Disk emitter model: the disk emitter model can explain the double-peaked BEL,
		but not the spatial separation of the double-peaked BEL.
		(c) SMBH--rSMBH model: a multiple SMBH system comprising SMBH1, SMBH2, and SMBH3 is shown.
		The merging of SMBH2 and SMBH3 yields an rSMBH with the emission of GWs.
		(d) Gravitational slingshot model: SMBH1 and SMBH2 constitute a hung-up bSMBH system,
		and SMBH3 is gravitationally ejected at a speed of several thousands of $\rm km\,s^{-1}$.
		The SMBH--rSMBH and gravitational slingshot models can explain
		the measured velocity offset and spatially separated distance of the double-peaked BEL of 1659$+$1834.
		\label{fig:Sch}}
\end{figure*}

 Several plausible models can be considered to explain the double-peaked BEL system:
(i) a bSMBH system; (ii) a disk emitter; (iii) an SMBH-rSMBH system; and (iv) a gravitational slingshot.
These possible systems are presented in Figure \ref{fig:Sch}.

 First, the bSMBH scenario is the most unlikely.
The secondary component does not have associated NELs,
and this disfavors the case of two AGNs brought to proximity recently ($\lesssim 1$\,Gyr).
The superposition of two AGNs in the same line of sight can be also excluded for the same reason.
Furthermore, if the separation is $\sim$250\,pc as suggested by some of the spaxel fitting solutions obtained in this study,
the bSMBH interpretation expects a velocity offset of $\sim$250\,$\rm km\,s^{-1}$
even if (i) the orbital eccentricity of the bSMBHs is 0.75 and
(ii) the BH in the secondary component is at perigee.
This is a factor of $\sim$10 smaller than the measured velocity offset of 3000\,$\rm km\,s^{-1}$. 

 Several AGNs exhibit double-peaked BELs,
and the disk emitter model can successfully explain such AGNs (e.g., \citealt{chen89a,chen89b}).
For the accretion disk of a $10^{9}\,M_{\rm \odot}$ black hole with
the disk extending out to $\sim$10 Schwarzschild radii \citep{morgan10},
the outer edge of the disk is $\sim0.002$\,pc.
This is far smaller than the size explored in this study,
and is consistent with some of our solutions for the spatial separation.
Therefore, the disk emitter model is viable.

 Another possible but exotic scenario is that 1695$+$1834 is a combined system comprising a primary SMBH and a secondary rSMBH.
rSMBHs can be recoiled at a speed up to several thousands of $\rm km\,s^{-1}$ \citep{campanelli07},
which matches the measured BEL velocity offset of 1695$+$1834.
Furthermore, BLRs could be gravitationally bound to rSMBHs;
however, rSMBHs leave narrow line regions behind to the host galaxy \citep{blecha11}.
This can explain why only the primary components of the BELs have the associated NELs.
Considering this scenario, there were three SMBHs originally, one in a galaxy and a close pair of SMBHs in another galaxy.
After these two galaxies merged, two active nuclei made of the three SMBHs came close to each other.
Then, the two close SMBHs merged and recoiled at a line-of-sight speed of $\rm {\sim 3000\,km\,s^{-1}}$.

 Finally, we also suggest a gravitational slingshot involving triple SMBHs
as a possible explanation \citep{haehnelt06}.
Three SMBHs are involved in this case as well.
The primary component is a hung-up binary SMBH system that is approached by a
relatively light SMBH with the BLR.
This light SMBH, that is, the secondary component, and its BLR can be ejected
at a speed of several thousands of $\rm km\,s^{-1}$ (e.g., \citealt{haehnelt06,hoffman06})
and identified as a spatially and kinematically offset AGN.

\section{Discussion} \label{section:discussion}
 In this study, we investigated the spatial distributions of the primary and secondary velocity components of
the double-peaked BEL of a red AGN of a merging galaxy, namely, 1659$+$1834.
The centroid offset between the two components is measured to be 0$\farcs$085
under several plausible velocity component fitting settings;
however, this result is ambiguous, as different answers,
including a near-null ($< 0\farcs05$) or a large spatial separation ($\sim0\farcs15$),
can be obtained if substantially extensive velocity component fitting settings are considered.
As there is no convincing evidence regarding the spatial separation of the BEL components,
several physical origins for the double-peaked emission lines of 1659$+$1834 could be possible,
including the origins explained via the disk emitter model and the more exotic rSMBH or slingshot SMBH models.

 Nevertheless, the possible separation of $0\farcs085$ that was obtained under several plausible BEL fitting assumptions in this study
favors the SMBH--rSMBH or gravitational slingshot model over the binary SMBH or disk emitter model.
This result encourages future research for finding more multiple SMBH systems in red AGNs.
High-angular-resolution observations,
such as those performed using the IFU with adaptive optics and VLBI in radio,
can possibly confirm the spatially separated nature of the BEL cores of 1659$+$1834
and detect more similar systems.
Our result also reveals interesting prospects regarding the electromagnetic (EM) counterpart search for GW events
triggered by binary BH mergers.
Several works suggest that EM emissions might occur owing to BH mergers in the accretion disks of AGNs (e.g., \citealt{bartos17}).
If the multiple SMBH scenario for 1659$+$1834 is accurate, red AGNs can be excellent candidates as
hosts of binary SMBH merger events that might be discovered through future missions such as the Laser Interferometer Space Antenna (LISA).

\acknowledgments

L.C.H. was supported by the National Science Foundation of China (11721303) and the National Key R\&D Program of China (2016YFA0400702).
D.K. acknowledges support from a KIAA Fellowship in addition to support from the China Postdoctoral Science Foundation (2019M650310).
This work was supported by the Creative Initiative Program of the National Research Foundation of Korea (NRF),
No. 2017R1A3A3001362, funded by the Korea government (MSIP).
The Gemini data were taken through the K-GMT Science Program
(PID: GN-2018A-Q-222) of Korea Astronomy and Space Science Institute (KASI).
PyRAF is a product of the Space Telescope Science Institute, which is operated by AURA for NASA.

\vspace{5mm}
\facilities{Gemini/North (GMOS/IFU)}
	
\software{
	IDL,
	L.A.Cosmic \citep{vandokkum01},
	\texttt{MPFIT} \citep{markwardt09},
	and PyRAF
}

\clearpage

\clearpage

\end{document}